# Proton LINAC Using Spiral Waveguide with Finite Energy of 80 MeV

*S. N. Dolya*

*Joint Institute for Nuclear Research, Joliot - Curie 6, Dubna, Russia, 141980*

**Abstract**

The article considers an opportunity of simultaneous pulsed acceleration of seven proton beams with current $I_p$ = 100 mA in each beam. The accelerator consists of two parts. In the first part of the accelerator having the length $L_{acc1}$ = 5 m, the protons are accelerated to the energy of $Є$ = 5 MeV. Consumption of high-frequency power by this part of the accelerator is equal to $P_1 \approx$ 10 MW. In the second part of the accelerator having the length $L_{acc2}$ = 50 m, the protons are accelerated to the finite energy $Є$ = 80 MeV. Consumption of the high frequency power by the second part of the accelerator is $P_2 \approx$ 74 MW. The radial focus of the proton beam in the first and second parts of the accelerator is carried out with the magnetic field H = 10 T which is generated by a superconducting solenoid.

### 1. Introduction

Article [1] considers the proton LINAC having energy $Є$ = 1 GeV and a pulse current of 0.7 A. The initial part of the accelerator is an assembly of the spiral waveguides mounted in a common screen. The finite energy acquired by protons in the initial part of the accelerator is $Є$ = 5 MeV, that corresponds to the finite velocity of the protons in the accelerator $β_{fin1}$ = 0.1, where β = v / c - the velocity of the protons, expressed in terms of the velocity of light, c = 3*10$^{10}$ cm / s.

It is assumed that the main part of the accelerator [1] consists of a sequence of superconducting cavities having resonance frequency $f_3$ = 1.3 GHz. This frequency corresponds to a wavelength $λ_3$ = c / $f_3$ = 23 cm. Assuming that the distance between the end walls of each cavity is equal to about βλ / 4, we see that at the beginning of the main part of the accelerator the distance between the end walls should be $β_{fin1}$ * $λ_3$ / 4 ≈ 0.6 cm. This distance (0.6 cm) between the end walls of each cavity seems to be too small for the beam acceleration.

In [1] it is proposed to connect all the seven beams into one by means of the doublet lenses having different focal lengths. As a result of this merger we obtain one beam with radius $r_b$ ≈ 1.5 cm. It is possible not to emerge the beams but simply make them close to each other and accelerate them individually, Fig 1. Then the radius of each beam ($r_{b\ ind}$ = 0.4 cm) will be much smaller than the radius of the beams joint together ($r_b$ ≈ 1.5 cm).

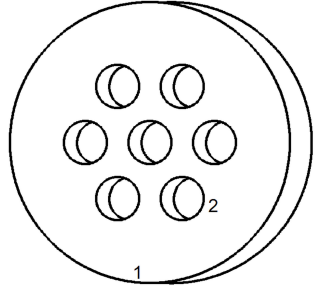

Fig. 1. 1 - Superconducting cylindrical cavity, 2 - holes through which the proton beams pass

We assume that each beam with radius $r_{b\ ind} = 0.4$ cm is accelerated by the cavities in their own aperture, Fig 1. We assume that the diameter of this aperture is equal to $d_{ap} = 15$ mm. Then the ratio of this diameter to the distance between the end walls of the cavity ( $l_{cav} = \beta_{fin1} * \lambda_3 / 4 \approx 0.6$ cm) is equal to $d_{ap} / l_{cav} = 2.5$, which is a too large value to accelerate the proton beam.

Note that in the initial part of the accelerator of the Alvarez type, where the initial velocity, expressed in terms of the velocity of light $\beta_{inA} = 0.04$, and the wavelength of acceleration $\lambda_A = 2$ m, the initial distance between the drift tubes is equal to $\beta_{inA} * \lambda_A / 4 = 2$ cm. To obtain this distance between the end walls of the cavity in the main part of the accelerator considered in [1], it is necessary to increase the initial velocity of the protons injected into the accelerator by about 4 times.

Indeed, in comparison with the accelerator of the Alvarez type which has frequency of proton acceleration $f_A = 150$ MHz, in the accelerator, discussed in [1], the frequency of proton acceleration is approximately by one order higher than $f_3 = 1.3$ GHz. To have the same value: $\beta_{inA} * \lambda_A / 4 = 2$ cm, as in the accelerator of the Alvarez type, the initial velocity of the protons injected into this part of the accelerator should have the value $\beta_{in3} = 0.4$. This is by an order of magnitude greater than the initial injection rate in the Alvarez accelerator and by 4 times more than the initial injection rate of the proton beam into the main part of the accelerator considered in [1].

The velocity $\beta_{in3} = 0.4$ corresponds to the energy of the protons $\epsilon_{fin2} = mc^2 \beta^2_{in3} / 2 = 80$ MeV. Thus, the accelerator, considered in [1], should be completed by an intermediate accelerator having the initial energy $\mathcal{E}_{in2} = 5$ MeV and finite energy $\mathcal{E}_{fin2} = 80$ MeV.

Consider an opportunity of creating such an accelerator based on the spiral waveguide.

**2. Selection of the main parameters of the accelerator**

We assume that the acceleration in the initial part of the LINAC considered in [1] is carried out at the frequency $f_1 = 260$ MHz, that is close to the acceleration frequency $f = 300$ MHz. Then, the acceleration in the main part of the accelerator will be carried out at the $5^{th}$ harmonic of frequency $f_1$, $f_3 = 5f_1 = 1.3$ GHz. In the intermediate part of the accelerator the acceleration will be carried out at the $3^{rd}$ harmonic of frequency $f_1$. Then the acceleration frequency in this part of the accelerator will be equal to $f_2 = 3f_1 = 780$ MHz.

*2.1. The choice of the initial part parameters of the accelerator*

For the initial part of the accelerator we have chosen the acceleration frequency $f_1 = 260$ MHz, that corresponds to the wavelength of the acceleration in vacuum $\lambda_1 = c / f_1 = 115$ cm. The radius of the beam and the beam injection voltage are remained the same, $r_b = 0.4$ cm, $U = 800$ keV, respectively. The choice of this injection voltage means that the initial velocity of the protons, expressed in terms of the velocity of light $\beta_{in1}$, will be equal to $\beta_{in1} = v_{in1} / c = 0.04$. The slow down wave length $\lambda_{1s}$ is equal to $\lambda_{1s} = \lambda_1 * \beta_{in1} = 4.6$ cm. Taking into account that the bunch takes over the phases, roughly, $\Delta\varphi = 36^0$, we find that the longitudinal dimension of the bunch is equal to $l_{bin1} = 4.6$ mm. Now you can calculate the volume of bunch: $V_{bin1} = \pi r_b^2 l_{bin1} = 0.23$ cm$^3$.

The current in each beam is assumed to be equal to $I_b = 100$ mA. Knowing the period of the wave with the frequency $f_1 = 260$ MHz, namely, $T_1 = 1 / f_1 = 3.85$ ns, we find that the number of protons in the bunch $N_b = I_b * T_1 = 6*10^{17} * 3.85*10^{-9} = 2.3*10^9$ protons in each bunch. Knowing the volume of the bunch $V_{bin1} = 0.23$ cm$^3$, it is possible to find the density of the particles in it: $n_{b1} = N_b / V_{bin1} = 10^{10}$ p / cm$^3$. The plasma frequency of the particle oscillations [1] in this bunch is as follows:
$\omega_{p1} = 1.3 * 10^3 * nb_1^{1/2} = 1.3 * 10^8$.

We assume that in the transverse direction the individual beams are held by the magnetic field $H = 10$ T, generated by the superconducting solenoid [1].

The corresponding Larmor frequency is equal to $\omega_L = eH / 2mc = 5 * 10^8$. It is larger than the plasma frequency and, to hold the size of the beam in the transverse direction will be possible. You just have to keep in mind that in the transverse direction to the action of the Coulomb forces of repulsion in this case there is the defocusing force acting on the protons from the side of the accelerating wave. Its value is estimate below.

We choose the amplitude of the accelerating field in the initial part of the accelerator to be equal to the following: $E_{01} = 20$ kV / cm, and the cosine of the synchronous phase $\cos\varphi_s = ½$. Thus, we have chosen the acceleration rate in the initial part of the accelerator to be equal to: $eE_{01} * \cos\varphi_s = 1$ MeV / m. To achieve the finite acceleration energy in the initial part of the accelerator $\mathcal{E}_1 = 5$ MeV, it is required to have the acceleration length $L_{acc1} = \mathcal{E}_1 / (eE_{01} * \cos\varphi_s) = 5$ m.

Let us find the value $W_\lambda$ equal to the ratio of the proton energy obtained at the wavelength to the rest energy of the proton ($mc^2 = 1$ GeV ), [2]:

$$W_\lambda = eE_0\lambda * \cos\varphi_s / mc^2. \qquad (1)$$

Substituting the numbers into the formula (1) we find that for the initial part of the accelerator the value $W_\lambda = eE_{01}\lambda * \cos\varphi_s / mc^2 = 1.15 * 10^{-3}$.

We find the frequency of the phase (longitudinal) oscillations of protons in the field of the accelerating wave [2]:

$$\Omega_{ph} = \omega*(W_\lambda*\tg \varphi_s/2\pi\beta_s)^{1/2}, \qquad (2)$$

where: $\omega = 2\pi f_1 = 1.63 * 10^9$ - circular frequency of the acceleration. Substituting the numbers into the formula (2) we find that in this case the frequency of the phase oscillations in the wave field is equal to: $\Omega_{ph1} = 1.43 * 10^8$. This frequency is higher than the plasma frequency: $\omega_{p1} = 1.3 * 10^3 * nb_1^{1/2} = 1.3 * 10^8$. It shows that the accelerating field of the wave will hold the longitudinal dimension of the bunch.

We now estimate the defocusing force acting on the protons from the side of the accelerating wave. The frequency $\Omega_{r1}$ [2] corresponding to this force is equal to $\Omega_r = \Omega_{ph} / 1.41$. In our case $\Omega_{r1} = \Omega_{ph1} / 1.41 = 10^8$, this frequency is of the order of the plasma frequency, and it is much smaller than the Larmor frequency.

We estimate the high-frequency power required to create the electric field tension $E_{01}$= 20 kV / cm in the spiral waveguide [3]:

$$P = (c/8)\, E_0^2\, r_{01in}^2\, \beta_{ph1}\{\ \},\qquad (3)$$

where $r_{01in}$ is the initial radius of the spiral. We remind that to maintain a uniform rate of acceleration along the accelerating section, the spiral should be wound on the narrowing cone. The expression in the curved brackets is a combination of Bessel functions of the first and second kinds. The numerical value of this expression is approximately equal to 8 [1], and this value slightly changes along the accelerating section. The initial radius of the spiral is chosen to be equal to $r_{01in}$ = 1 cm. Substituting the numbers into the formula (3) we get the following:

$$P\,(W) = 3*10^{10}*4*10^{8}*4*10^{-2}*8/(8*9*10^{4}*10^{7}) = 0.53\ \text{MW}.$$

We must also take into account that about the same power of $\mathrm{\mathcal{E}} * I_b = 0.5$ MW is transmitted from the high-frequency electric field to each beam, so that the total power transmitted in each beam is approximately 1 MW. Since the accelerator contains seven beams, the total high-frequency power consuming by the first part of the accelerator, will be 7 MW.

Here is the Table of parameters of the initial part of the accelerator.

Table1. Parameters of the initial part of the accelerator

| Option | Value |
| --- | --- |
| The voltage of the proton source is, kV | 800 |
| Bunched beam current, mA | 7*100 |
| The frequency of the acceleration f, MHz | 260 |
| The tension of the focusing magnetic field, T | 10 |
| The duration of the current pulse $\tau_b$, μs | 200 |
| The average tension of electric field $E_0$, MV /m | 2 |
| The cosine of the synchronous phase, $\cos\varphi_s$ | 0.5 |
| The pulse repetition rate F, Hz | 5 |
| High-frequency power, MW | 7 |
| The length of the accelerator, m | 5 |

## 2. 2. The choice of parameters of the main accelerator part

The initial radius of the spiral will remain the same: $r_{0in2} = 1$ cm, towards the end of the accelerator, the radius must decrease [1] to roughly preserve constant rate of acceleration along all the length of the accelerator.

In this case the value of the dimensionless parameter x which is the argument of the Bessel functions will be equal to $x_{in} = 2\pi r_{0in2} / \beta_{in2}\lambda_2 = 1.63$, where we have taken $\lambda_2 = c / f_2 = 38.46$ cm that is the wavelength of acceleration. The finite value of parameter $x_{fin}$ will be by 4 times less than the initial one ($x_{fin} = 0.4$) that will not lead to a strong decrease of the amplitude of the accelerating field $E_{02}$ in this part of the accelerator [1].

We choose the tension amplitude of the accelerating field for the second part of the accelerator to be equal to $E_{02} = 30$ kV / cm, the cosine of the synchronous phase is left the same: $\cos \varphi_s = ½$. Then the acceleration energy rate in this part of the accelerator is equal to $eE_{02} * \cos \varphi_s = 1.5$ MeV / m and 75 MeV will be obtained by protons over length $L_{acc2} = 75 / 1.5 = 50$ m.

Let us find value $W_\lambda$ which is the value of energy acquired at the wavelength $\lambda$, referred to the rest mass of the proton equal to $mc^2 = 1$ GeV. Substituting the numbers into the formula (1) we find that in this case:
$W_\lambda = eE_{02}\lambda * \cos\varphi_s / mc^2 = 5.77 * 10^{-4}$.

The frequency of the phase oscillations of protons in the accelerating wave field [2] for the second part of the accelerator (formula 2) is equal to
$\Omega_{ph2} = \omega * (W_\lambda * \text{tg } \varphi_s / 2\pi\beta_s)^{1/2} = 1.96 * 10^8$, where $\omega = 2\pi f_2 = 4.9 * 10^9$ is the angular frequency of acceleration. This frequency must be compared with the plasma frequency.

We assume that the beam radius while transferring from the first to the second part of the accelerator has not changed. As for the longitudinal dimension of the bunch, it has, on the one hand, stretched by 2.5 times in accordance with the increase of the velocity of the beam from $\beta_{in1} = 0.04$ to $\beta_{fin1} = 0.1$. On the other hand, due to the transfer to frequency $f_2$, which is by three times greater than $f_1$, the longitudinal dimension of the bunch will reduce by 3 times. This will result in increasing the density of protons in the bunch in the second part of the accelerator by 20% in comparison with the first part of the accelerator. Thus, it will result in the increase of the plasma frequency by 10% till the value of $\omega_{p2} = 1.43 * 10^8$.

The frequency of the radial oscillations associated with proton defocusing caused by the accelerating wave field is as follows: $\Omega_{r2} = \Omega_{ph2} / 1.41 = 1.4 * 10^8$. It is approximately equal to the plasma frequency, and each of these both frequencies is much smaller than the Larmor frequency $\omega_L = 5 * 10^8$. This means that the magnetic field of 10 T, generated by the superconducting solenoid will keep the radial dimension of the bunch.

Now we determine the high frequency power required to generate the electric field tension $E_{02} = 30$ kV / cm in the second part of the accelerator. Substituting the numbers into formula (3) we find this power:

$$P (W) = 3*10^{10}*9*10^8*0.1*8/(8*9*10^4*10^7) = 3 \text{ MW}.$$

The beam will then obtain the power $P_b = \Delta \mathcal{E} * I_b = 75*0.1 = 7.5$ MW. The total consumption (by seven beams) of the high-frequency power is $P = 7*(3 + 7.5) = 74$ MW.

Here is the Table of parameters of the second part of the accelerator.

Table 2. Parameters of the second part of the accelerator

| Option | Value |
|---|---|
| Initial energy of the proton in MeV | 5 |
| Bunched beam current, mA | 7*100 |
| The frequency of the acceleration f, MHz | 780 |
| The tension of the focusing magnetic field, T | 10 |
| The duration of the current pulse $\tau_b$, μs | 200 |
| The average tension of electric field $E_0$, MV /m | 3 |
| The cosine of the synchronous phase, $\cos\varphi_s$ | 0.5 |
| The pulse repetition rate F, Hz | 5 |
| High-frequency power, MW | 74 |
| The length of the accelerator, m | 50 |

**3. Conclusion**

From the presented above it is clear that by means of the spiral waveguide it is possible to accelerate proton beams of high intensity (7 * 100 mA) till the intermediate energy $\mathcal{E} = 80$ MeV. This spiral waveguide has a rather high efficiency of acceleration, i.e., the total length of the accelerator will not exceed $L_{acc} < 60$ m.